\documentstyle[psfig,12pt]{article}

\renewcommand{\baselinestretch}{2}

\begin{document}

\title{New Reflections on Electron's Energy and Wavefunction in the Hydrogen Atom}

\renewcommand{\baselinestretch}{1}

\author{
{\large Ezzat G. Bakhoum}\\
{University of West Florida}\\
{11000 University Parkway, Pensacola, Fl. 32514 USA}\\
{Email: ebakhoum@uwf.edu}\\
{Copyright \copyright 2008-2009 by Ezzat G. Bakhoum}
}

\maketitle

\begin{center}
{\large\bf Abstract}\\
\end{center}

Schr\"{o}dinger's equation predicts something very peculiar about the electron in the Hydrogen atom: its total energy must be equal to zero. Unfortunately, an analysis of a zero-energy wavefunction for the electron in the Hydrogen atom has not been attempted in the published literature. This paper provides such an analysis for the first time and uncovers a few interesting facts, including the fact that a ``zero-energy wavefunction" is actually a quantized version of the classical wavefunction that has been known for decades.

\newpage

{\large\bf 1. Introduction:}

The principle of mass-energy equivalence normally never comes to mind when the quantum-mechanical analysis of the hydrogen atom is undertaken. It is well known that by applying Schr\"{o}dinger's equation to the problem of the electron in the hydrogen atom, the Balmer energy levels are obtained by means of a purely classical (i.e., non relativistic) analysis \cite{Griffiths, French}.  There is, however, a very interesting connection between the problem of the hydrogen atom and the principle of mass-energy equivalence that was previously unexplored. If we write the time-dependent Schr\"{o}dinger equation 
\begin{equation}
i \hbar \frac{\partial\psi}{\partial t} = H \; \psi
\label{1}
\end{equation}

and its solution
\begin{equation}
\psi = \psi_0 \; \exp \left( - \frac{i}{\hbar} \int H dt \right),
\label{2}
\end{equation}

where $H$ is the total energy of the moving particle, namely, the electron; we must ask what conclusion can we make if we assume that the electron is in a stable orbit around the nucleus? Obviously, we must assume that the wave function $\psi=\psi_0$ (i.e., a constant, or stable wave function that does not evolve over time). This, of course, is well known, since the electron's wave function in the hydrogen atom represents a standing wave and has no time dependence. Hence, the conclusion that inevitably emerges in this case is that the total energy of the electron $H$ must be equal to zero everywhere along the path of the electron. In view of some fundamental research on the principle of mass-energy equivalence that was previously published by the author \cite{Bakhoum1, Bakhoum2}, this conclusion, as a matter of fact, is not surprising.\\
\\
In the earlier publications by the author, it was demonstrated that a number of fundamental problems in quantum mechanics cannot be understood on the basis of the relativistic law of mass-energy equivalence, $H = m c^2$. The problem of the hydrogen atom is one such problem. It was further demonstrated that $H = m c^2$ can be regarded as a {\em special case} of a more general law of mass-energy equivalence that does in fact explain that category of problems that the relativistic law fails to explain. That general law is $H = m v^2$, where the relativistic constant $c^2$ has been replaced by $v^2$, $v$ being the velocity of the moving particle (see references \cite{Bakhoum1, Bakhoum2, Bakhoum3} for a complete historical accounting of the origin and the applications of that law). We shall now proceed to solve the problem of the total energy of the electron in the hydrogen atom and demonstrate that the general mass-energy equivalence law $H = m v^2$ correlates with and explains the result predicted by Schr\"{o}dinger's equation. We shall further demonstrate that a new ``zero-energy wavefunction" that will be obtained under that law is actually a quantized version of the classical wavefunction that has been known for decades.\\
\\
{\large\bf 2. The law of mass-energy equivalence, and the ``zero-energy" wave equation:}

It is not difficult to see how the general mass-energy equivalence law $H = m v^2$ (which, admittedly, may seem strange to the readers who are not familiar with it) correlates with the result predicted by Schr\"{o}dinger's equation. In the hydrogen atom, the electron is in equilibrium due to the equality of the two forces
\begin{equation}
\frac{e^2}{r^2} = \frac{mv^2}{r}
\end{equation}

where $e^2/r^2$ is the Coulomb electrostatic force (here, $e^2 = q^2 /4 \pi\epsilon_0$, where $q$ is the electron's charge), and where $mv^2/r$ is the centrifugal force. But the electrostatic potential $V$ acting on the electron is equal to $-e^2/r$, by definition. From the above equation, it is therefore clear that $V = - mv^2$. If we now assume that the total energy of the free electron is given by the quantity $+ mv^2$, then it must be further clear that the total energy of the bound electron must be equal to zero (due to the addition of the electrostatic potential $V$)\footnote{It is to be pointed out that this conclusion concerns the TOTAL ENERGY of the electron. In practice, the atom is observed to emit and absorb energy during bound-state transitions because such transitions involve only kinetic energy and potential energy changes. Mass-energy equivalence obviously does not play a role in electronic bound-state transitions. That is why the present conclusions are not in disagreement with the classical theory or with experimental results.}. This is the classical view according to Bohr's theory. Let us now examine the view according to the Schr\"{o}dinger Hamiltonian theory.\\
\\
The classical Schr\"{o}dinger Hamiltonian is given by
\begin{equation}
H = - \frac{\hbar^2}{2 m} \nabla^2 + V
\label{3}
\end{equation}

This Hamiltonian represents the sum Kinetic Energy + Potential Energy, and it is the Hamiltonian used to derive the Balmer energy levels and the classical wave function of the electron. If we want to write the Hamiltonian in a manner that takes mass-energy equivalence into account, the Hamiltonian will be written as follows:
\begin{equation}
H = - \frac{\hbar^2}{m} \nabla^2 + V,
\label{4}
\end{equation}

where we have replaced the kinetic energy $1/2 \; mv^2$ by the total energy $mv^2$. But since the total energy must be equal to zero, then we have the following wave equation
\begin{equation}
- \frac{\hbar^2}{m} \nabla^2 \psi_0 + V \psi_0 = 0
\label{5}
\end{equation}

We shall now demonstrate that the wave function $\psi_0$ that satisfies this zero-energy wave equation is the same as the wave function derived through the classical analysis, with the surprising restriction that the wave function itself must be radially quantized!\\
\\
{\large\bf 3. The connection between the zero-energy wave equation and the classical wave equation:}
\\
For the purpose of comparison, we write the classical equation that is based on the Schr\"{o}dinger Hamiltonian together with the new wave equation that incorporates mass-energy equivalence:

\begin{equation}
\begin{array}{rcll}
- (\hbar^2/2m) \; \nabla^2 \psi_0 + V \psi_0 & = & W \: \psi_0 \qquad & \mbox{(classical)} \\
- (\hbar^2/m) \; \nabla^2 \psi_0 + V \psi_0  & = & 0           \qquad & \mbox{(total energy)}
\end{array}
\label{6}
\end{equation}

where $W$ represents the Balmer energy levels and where $V = -e^2/r$ is the potential of the nucleus. While the two equations obviously seem to be two very different equations,  we shall now demonstrate that the second equation does {\em indeed} revert to the first equation if $\psi_0$ is restricted to be a radially quantized function, rather than a continuous function! We first write the zero-energy equation as follows:
\begin{equation}
(\hbar^2/m) \; \nabla^2 \psi_0 = V \psi_0  = - \frac{e^2}{r} \; \psi_0
\label{7}
\end{equation}

Since the radial distance $r$ takes only quantized values as multiples of the Bohr radius, $a = \hbar^2 / m e^2$, we substitute for $r$ in the equation by using this quantity, getting,
\begin{equation}
(\hbar^2/m) \; \nabla^2 \psi_0 = - e^2 \; \frac{m e^2}{\hbar^2} \; \psi_0 = - \frac{m e^4}{\hbar^2} \; \psi_0
\label{8}
\end{equation}

Dividing both sides of the equation by 2 gives
\begin{equation}
(\hbar^2/2m) \; \nabla^2 \psi_0 = - \frac{m e^4}{2 \hbar^2} \; \psi_0
\label{9}
\end{equation}

It is not difficult to verify that the coefficient of $\psi_0$ on the r.h.s. of the equation is the Balmer energy $W$. That is, we have the result that
\begin{equation}
(\hbar^2/2m) \; \nabla^2 \psi_0 = W \; \psi_0
\label{10}
\end{equation}

Now, by virtue of Eq.(\ref{10}), the zero-energy wave equation in (\ref{6}) can be finally written as
\begin{eqnarray}
- (\hbar^2/2m) \; \nabla^2 \psi_0 + V \psi_0  & = & + (\hbar^2/2m) \; \nabla^2 \psi_0 \nonumber\\
                                             & = & W \; \psi_0
\label{11}
\end{eqnarray}

This last equation is of course the classical wave equation.\\
\\
If we decompose the zero-energy wave equation into its radial and spherical-harmonic components, it becomes a simple matter to verify that the classical unnormalized wave function 
\begin{equation}
\psi_0(r) = \exp \left( - \frac{me^2}{\hbar^2} \: r \right)
\label{12}
\end{equation}

will indeed satisfy the radial wave equation at $r = na$, or integer multiples of the Bohr radius (see the proof in the Appendix). The fact that the classical wave function satisfies the zero-energy wave equation at multiples of the Bohr radius can be understood physically as follows: the classical wave function is a continuous, differentiable function that defines the boundary of a space that theoretically extends from $r=0$ to $r=\infty$ (see the plot in Fig. 1). The solution of the zero-energy wave equation, on the other hand, is a discrete, sparse set in $r$ that is defined {\em only} at integer multiples of the Bohr radius (see figure). This discrete function therefore inhabits the space defined by the classical wave function (a simple analogy might be a wave in a plastic sheet on top of which tiny droplets of mercury always flow to the minimum of that wave, as if the wave was a ``potential well"). This is not a surprise, since, as was concluded earlier, the total energy of the electron is equal to zero at multiples of the Bohr radius. The discrete solution, therefore, is indeed a solution in which the minimum energy principle is manifested; as opposed to the classical solution in which only the kinetic and potential energies are accounted for.

\newpage

\centerline{\psfig{figure=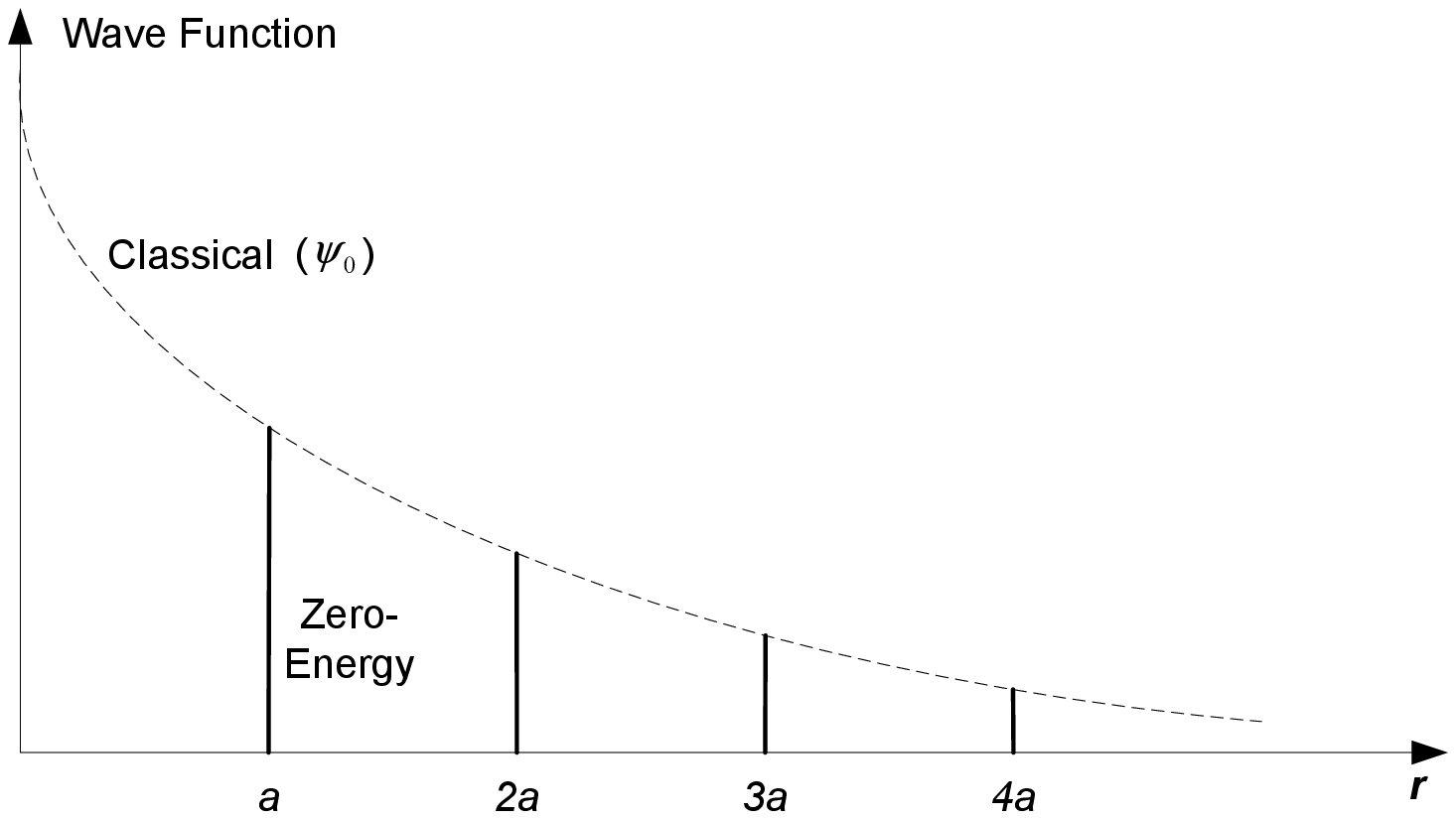,width=4in}}

\vspace*{.1in}

Figure 1: The classical wave function $\psi_0$ and the quantized solution of the zero-energy wave equation. The latter exists only at integer multiples of the Bohr radius and inhabits the space defined by the former.\\
\\
{\large\bf Appendix: Solution of the zero-energy wave equation, and the quantization condition}

To solve Eq.(\ref{5}) for $\psi_0$, we must replace the operator $\nabla^2$ by its equivalent expression in spherical coordinates, and substitute for the potential $V$ by  the traditional quantity $-e^2/r$. The process of replacing $\nabla^2$ in Eq.(\ref{5}) by its equivalent expression in spherical coordinates is well known in the literature \cite{Griffiths, French}, and we simply write the result:

\begin{equation}
\left( - \frac{\hbar^2}{m} \frac{d^2}{dr^2} + \frac{l(l+1)\hbar^2}{mr^2} - \frac{e^2}{r} \right) \left( r \psi_0(r) \right) = 0
\label{a1}
\end{equation}

Here, $\psi_0(r)$ is the radial component of $\psi_0$ and $l$ is the orbital quantum number. Typically, a second equation is needed to solve for the spherical-harmonic component of $\psi_0$, but since this solution is well known in the literature it will not be discussed here. The usual approach for solving Eq.(\ref{a1}) is to let the product $r \psi_0(r)$ be equal to another function, say $\Gamma(r)$. Eq.(\ref{a1}) is then rewritten as

\begin{equation}
- \Gamma^{\prime\prime} (r) + \left( \frac{l(l+1)}{r^2} - \frac{me^2}{\hbar^2} \cdot \frac{1}{r} \right) \; \Gamma(r) = 0
\label{a2}
\end{equation}

In the classical solution, the Balmer series for hydrogen is obtained by simply setting $l = 0$. While the above equation cannot be solved for the Balmer energy, setting $l = 0$ results in

\begin{equation}
\Gamma^{\prime\prime} (r) + \frac{me^2}{\hbar^2} \cdot \frac{1}{r} \; \Gamma(r) = 0
\label{a3}
\end{equation}

We now note that the quantity $\hbar^2 / me^2$ represents the Bohr radius, $a$. We shall follow however the standard procedure of replacing $a$ by $na$, where $n$ is the principal quantum number. We therefore rewrite the above equation as follows:

\begin{equation}
\Gamma^{\prime\prime} (r) + \frac{1}{na} \cdot \frac{1}{r} \; \Gamma(r) = 0
\label{a4}
\end{equation}

Solving this simple differential equation is a simple but rather lengthy and uninformative mathematical exercise. It can be quickly verified, however, that the classical wave function

\begin{equation}
\Gamma(r) = r \: \psi_0(r) = r \: \exp \left( - \frac{r}{na} \right)
\label{a5}
\end{equation}

does in fact satisfy Eq.(\ref{a4}), provided that the radial distance $r$ in the final expression is replaced by an integer multiple of the Bohr radius, or $na$.\\
\\
{\large\bf Acknowledgments:}

The author is highly indebted to Dr. Michael Brill, Associate Editor of Physics Essays, for suggesting the topic of this paper and for numerous valuable discussions and suggestions.


\vspace*{.5in}


\begin{thebibliography}{99}

\bibitem{Griffiths} D. Griffiths, {\sl Introduction to Elementary Particles\/} (John Wiley, New York, NY, 1987).

\bibitem{French} A.P. French and E.F. Taylor, {\sl An Introduction to Quantum Physics\/} (Norton Publications, New York, NY, 1978).

\bibitem{Bakhoum1} E. Bakhoum, {\sl Fundamental Disagreement of Wave Mechanics with Relativity\/}, Physics Essays, 15, 1, 2002. Online e-print archive: physics/0206061.

\bibitem{Bakhoum2} E. Bakhoum, {\sl On the Equation $H=mv^2$ and the Fine Structure of the Hydrogen Atom\/}, Physics Essays, 15, 4, 2002. Online e-print archive: physics/0207107.

\bibitem{Bakhoum3} E. Bakhoum, {\sl Electrodynamics and the Mass-Energy Equivalence Principle\/}, Physics Essays, 19, 3, 2006. Online e-print archive: physics/0310019.



\end{thebibliography}
\end{document}